\documentstyle[aps,preprint]{revtex}
\tightenlines
\begin{document}
\def\ii{\'{\i}}
\def\bi{\bigskip}
\def\be{\begin{equation}}
\def\en{\end{equation}}
\def\bq{\begin{eqnarray}}
\def\eq{\end{eqnarray}}
\def\noi{\noindent}
\def\bc{\begin{center}}
\def\ec{\end{center}}
\def\beit{\begin{itemize}}
\def\eit{\end{itemize}}
\def\ct{\centerline}
\def\tr{{\rm tr}}

\title{``Principle of Indistinguishability'' and equations of motion for 
particles with spin \footnote{Invited talk at Zacatecas Forum in Physics, 
Zacatecas, Mexico, May 11-13 2002.} }
\author{ Mauro Napsuciale \footnote{email:mauro@feynman.ugto.mx} \\
{\small\it Instituto de Fisica, Universidad de Guanajuato,\\
Lomas del Bosque 103, Fracc. Lomas del Campestre \\ 
37150, Leon, Guanajuato, Mexico }}
\maketitle

\begin{abstract}
In this work we review the derivation of Dirac and Weinberg equations based 
on a ``principle of indistinguishability'' for the $(j,0)$ and $(0,j)$ 
irreducible representations (irreps) of the Homogeneous Lorentz Group 
(HLG). We generalize this principle and explore its consequences for other 
irreps containing $j\ge 1$. We rederive Ahluwalia-Kirchbach equation using 
this principle and conclude that it yields ${\cal O}(p^{2j})$ equations of 
motion for any representation containing spin $j$ and lower spins.
We also use the obtained generators of the HLG for a given representation
to explore the possibility of the existence of first order equations for 
that representation. We show that, except for $j=\frac{1}{2}$, there exists no 
Dirac-like equation for the $(j,0)\oplus (0,j)$ representation nor for the 
$(\frac{1}{2},\frac{1}{2})$ representation. We rederive 
Kemmer-Duffin-Petieau (KDP) equation for the 
$(1,0)\oplus (\frac{1}{2},\frac{1}{2}) 
\oplus(0,1)$ representation by this method and show that the 
$(1,\frac{1}{2})\oplus (\frac{1}{2},1)$ representation satisfies a 
Dirac-like equation which describes a multiplet of $j=\frac{3}{2}$ and 
$j=\frac{1}{2}$ with masses $m$ and $\frac{m}{2}$ respectively.  
\end{abstract}

\bi

\section{Introduction}

The field theoretical description of interactions of particles with spin 
$> 1$  is a long standing problem. The interaction of a spin 
$\frac{3}{2}$ Rarita-Schwinger (RS) field minimally coupled to an external 
electromagnetic field was shown
to be inconsistent more than fourthy years ago \cite{sudarshan1}. After 
this seminal work many authors have addressed this problem 
from different perspectives and for different interactions \cite{todos}.
In particular the frequently posed requirement on unphysicality
of the spin $\frac{1}{2}$ content of the RS field is the source
of ambiguities in  the description of the interactions of spin 3/2 
particles with external fields (the so-called ``off-shell'' ambiguities\cite{offshell}).
The conclusion seems to be that it is not possible to construct a quantum
theory of higher spin interacting particles and even some no-go theorems 
have been formulated \cite{witten} for the existence of massless particles 
with spin $>1$. From a 
very phenomenological perspective this conclusion is very disappointing since
 on the one hand it closes the door for the possible existence of fundamental 
particles with $s>1$ and on the other hand there exist a plethora of 
resonances with $s>1$. Certainly, we know these are composite particles but 
in the long wavelength regime their composite nature is completely 
irrelevant and manifest only in the values of a few parameters 
(low energy constants). Thus, the long wave description of composite 
particles calls for a description for elementary systems with 
the any spin. This is particularly relevant for effective field 
theories of hadrons. In 
particular, the systematic expansion in powers of the momentum and quark 
masses for pseudoscalars \cite{chpt}, the so-called chiral perturbation theory 
(CHPT), is lost when we incorporate spin $\frac{1}{2}$ baryons \cite{bchpt}. 
Recovering a systematic expansion requires a heavy field expansion 
 for spin $\frac{1}{2}$ degrees of freedom \cite{hbchpt}. In this framework, 
the spin $\frac{3}{2}$ degrees of freedom have been shown to play a 
prominent role since  they are not heavy enough to be ``integrated out'' 
and these degrees of freedom must be considered from the very start \cite{jenkins}. 
Ordinarily, they are treated within the frame work
of the Rarita-Schwinger (RS) formalism. Clearly, one has to be
careful when considering degrees of freedom with $s>1$ because
of fundamental quantum inconsistencies in the description of their 
interactions. However, as we showed in Ref.~\cite{napsu1},
the leading order of the heavy field expansion  is free of 
the ``off-shell'' ambiguities and quantum inconsistencies. 
Beyond this order one  cannot avoid  problems when using the RS formalism 
for the description of spin $\frac{3}{2}$ degrees of freedom. 

History has given plenty of examples that looking
on phenomena from a perspective different but the 
originally accepted one, can lead to new and even surprising 
insights and in additon bring technical advantages. Historically 
the discovery of the equation of motion for spin $\frac{1}{2}$ particles by 
Dirac in 1931 was motivated by the desire to find a resolution
of the problem of negative probabilities of the Klein-Gordon 
equation. After seven decades we are facing many different approaches to 
Dirac's equation. In one of the possibilities,
it can be viewed just as a consequence of the transformation 
properties of the corresponding creation and annihilation operators 
under the Poincare Group \cite{weinberg,weinbergbook} 
A different approach was put forward by Ryder in his well known
textbook \cite{ryder}. Ryder's method is based upon the 
representation theory of the Homogeneous Lorentz Group (HLG) 
in combination with certain identities valid only in the rest frame
for which we here coined the term  
``{\it principle of indistinguishability\/}''. 
It is that very principle on which we shall focus in the following.

In Ref.~\cite{ahlu1}, Ryder's method was extended to 
incorporate discrete, $C$, $P$, and $T$ symmetries into
the wave equations for $(j,0)\oplus(0,j)$ by means of relative
phases between $(j,0)$, and $(0,j)$.
In particular Weinberg equations \cite{weinberg} have been
shown to follow from  specific choices for the corresponding 
phases. 
The obtained equations of motion are of the order
${\cal O} (p^{2j})$ in the momenta.  Hence, 
for $s>1$, the above procedure yields equations which are 
${\cal O}(p^3)$ or higher order in $p$.
The latter are known to possess acausal energy-momentum dispersion relations
beyond the standard ones.

\noindent
It is the goal of the present paper to  generalize the 
``principle of indistinguishability''  to representations containing 
$s>1$ and other but $(j,0)\oplus (0,j)$. 

We also constructed generators for a 
representation of interest from the generators of the simplest 
representations $(j,0)$ and $(0,j^\prime )$.
Within this scheme we explore possibility for Dirac-like 
equations to exist for arbitrary spin.

\section{Homogeneous Lorentz Group and principle of indistinguishability 
for the $(j,0)\oplus (0,j)$ representations: brief review}

\subsection{ Irreducible representations}
The local structure of the Homogeneous Lorentz group comprising 
boosts and rotations is expressed by means of the following
commutators

\begin{equation}
\left[ J_{i},J_{j}\right] =i\varepsilon _{ijk}J_{k}, ~~ 
\left[ J_{i},K_{j}\right] =i\varepsilon _{ijk}K_{k}, ~~ 
\left[ K_{i},K_{j}\right] =-i\varepsilon _{ijk}J_{k}. \label{algebraJK}
\nonumber
\end{equation}

\noi The  relations can be rewritten in terms of the new generators 
$\vec{A}=\frac{1}{2} \left( \vec{J}+i\vec{K}\right)$ and 
$\vec{B}=\frac{1}{2} \left( \vec{J}-i\vec{K}\right)$ which satisfy  

\begin{equation}
\left[ A_{i},A_{j}\right] =i\varepsilon _{ijk}A_{k}, ~~
\left[ B_{i},B_{j}\right] =i\varepsilon _{ijk}B_{k}, ~~ 
\left[ A_{i},B_{j}\right] =0.\label{algebraAB}
\end{equation}
\noi Equations (\ref{algebraAB}) shows that HLG is locally isomorphic 
to $SU(2)_A\otimes SU(2)_B$. 
This facilitates the classification of the irreps for 
this group which can be induced from those of $SU(2)$. Indeed, quantum states 
with well defined transformation properties under HLG can be classified 
according to  two ``angular momentum'' labels $(j,j^{\prime})$ corresponding 
to the two subgroups spanned by the generators $\vec A$ and $\vec B$. 
Furthermore, under parity $\vec J\to\vec J$, $\vec K \to - \vec K$, thus  
$\vec A \to\vec B$, $\vec B\to \vec A$, hence the $( j,j')$ and $(j',j)$ 
representations are interchanged under parity. Under a Lorentz  transformation
spinors in the $(j,j' )$ representation (in momentum space) transform as

\begin{equation}
\Phi \left( p^{\mu }\right) =\Lambda \left( 
\dot p^{\mu }\to p^{\mu }\right)\Phi \left( \dot p^{\mu }\right) 
=e^{i(\vec{J}\cdot  
\vec{\vartheta}+\vec{K}\cdot \vec{\varphi} )}
~\Phi \left( \dot p^{\mu }\right)\, ,  
\end{equation}
\noi where $\vec\vartheta ,\vec\varphi $ denote the parameters of the 
transformation.

\subsection{$(j,0)$ and $(0,j)$  representations.}

The $(j,0)$ and $(0,j)$ are the simplest irreps of the HLG. For the 
$ \left( j,0\right)$ 
representation we have $\vec{B}=0$, i.e. $\vec{J}=i\vec{K} $. For the 
$ \left( 0,j\right) $ representation $\vec{A}=0$ implies $\vec{J}=-i\vec{K}$.
Following the literature we denote states belonging to  the $(j,0)$ 
representation 
as "right" states  and those  in the $(0,j)$ representation
 as``left'' states \footnote{This nomenclature comes from the fact that in the massless case the 
``left'' and ``right'' spinors turn out to be eigenstates of helicity operator 
$\vec J\cdot \hat p$ with eigenvalues $-s$ (``left-handed'') and $+s$ 
(``right-handed'') respectively. In the massive case however is just a book-keeping 
device for the site where the $0$ is located in the $(j,0)$ or $(0,j)$ notation.}.
Under an HLG transformation, left and right states in momentum 
space transforms as
\begin{equation}
\Phi _{R}\left( p^{\mu }\right) =\Lambda _{R}\left( 
\dot p^{\mu }\to p^{\mu }\right)\Phi _{R}\left( 
\dot p^{\mu }\right) =e^{ i\vec{J}\cdot ( 
\vec{\vartheta}-i \vec{\varphi} )}~\Phi _{R}\left( 
\dot p^{\mu }\right), \label{hltr} 
\end{equation}
\begin{equation}
\Phi _{L}\left( p^{\mu }\right) =\Lambda _{L}\left( \dot p^{\mu }\to 
p^{\mu }\right)\Phi _{L}\left( 
\dot p^{\mu }\right) =e^{i\vec{J}\cdot (
\vec{\vartheta }+i\vec{\varphi })}~\Phi _{L}\left( 
\dot p^{\mu }\right). \label{hltl}
\end{equation}
\noi Pure Lorentz transformations (boosts) are obtained setting 
$\vec\vartheta=0$ in(\ref{hltr},\ref{hltl}) 
\begin{equation}
\Phi _{R}\left( p^{\mu }\right) =B _{R}\left( \dot p^{\mu }
 \to p^{\mu }\right))\Phi _{R}\left( \dot p^{\mu }\right) 
 =\exp (+\vec{J}\cdot \vec{\varphi })\Phi _{R}\left( \dot p^{\mu }\right),  
\label{boostR}
\end{equation}
\begin{equation}
\Phi _{L}\left( p^{\mu }\right) = B_{L}\left( \dot p^{\mu }
\to p^{\mu }\right)\Phi _{L}\left( \dot p^{\mu }\right)
 =\exp (-\vec{J}\cdot \vec{\varphi })\Phi _{L}\left( \dot p^{\mu }\right).  
\label{boostL}
\end{equation}
\noi In the case $\dot p^\mu =(m,\vec 0) $, where $m$ stands for the mass of 
the corresponding particle (i.e. when $\dot p^\mu$ is the momentum in the 
rest frame of the particle), the angular momentum generators are just the spin 
operators. In this case, the parameters $\vec{\varphi }$ are related to the energy 
and momentum of the particle in the boosted frame as follows
\begin{equation}
\cosh \varphi =\gamma =\frac{E}{m}\quad \sinh
\varphi =\beta \gamma =\frac{\left| \vec{p}\right| }{m}\quad 
\widehat{\varphi }=\frac{\vec{p}}{\left| 
\vec{p}\right| }. \label{phi(E,p)}
\end{equation}
\noi Under rotations ($\vec\varphi =0$ in (\ref{hltr},\ref{hltl})) 
the left and right states transform as 
\begin{equation}
\Phi _{R}\left( p^{\mu }\right) =R_{R}\left( \dot p^{\mu }
 \to p^{\mu }\right))\Phi _{R}\left( \dot p^{\mu }\right) 
 =\exp (i\vec{J}\cdot \vec{\vartheta })
\Phi _{R}\left( \dot p^{\mu }\right)\, , 
\label{rotR}
\end{equation}
\begin{equation}
\Phi _{L}\left( p^{\mu }\right) = R_{L}\left( \dot p^{\mu }
\to p^{\mu }\right)\Phi _{L}\left( \dot p^{\mu }\right)
 =\exp (i\vec{J}\cdot \vec{\vartheta })\Phi _{L}
\left( \dot p^{\mu }\right)\, .  
\label{rotL}
\end{equation}
Notice that rest frame states in the $(j,0)$ representation have 
exactly the same transformation properties under rotations than 
those in the $(0,j)$ representation. Thus  if particles are to be 
identified with the irreps of 
the HLG we are lead to the conclusion that there exist two kinds of states 
which are distinguished by their transformation properties under boosts but 
which when posed in the rest frame cannot be distinguished by their 
transformation properties under rotations. This is what we call the  
\bc
{\bf Principle of Indistinguishability (PI) }
\ec

{\it At rest, a spinor belonging to the $(j,0)$ representation is indistinguishable from
from a  spinor in the $(0,j)$ representation }\footnote{As far as I know this principle was first formulated 
for the case $j=\frac{1}{2}$ in 
the first edition of Ryder's book   \cite{ryder} although incomplete due to the 
missing of the phase $\rho$. This phase and its relation to intrinsic 
parity and anti-particle solutions to Dirac equation were firstly noticed in 
Refs.~\cite{ahlu1,argentinos}. 
Further intriguing work on the consequences of  
relative phases between the building blocks of composite
representations has been done in Refs.~\cite{ahlu1,ahlu2}.}.
 Thus, the corresponding quantum states can differ at most by a 
phase \footnote{ Hereafter we will use the tri-momentum as the argument of spinors.}  

\begin{equation}
\Phi _{R}(\vec 0) =\varrho ~\Phi _{L}(\vec 0),  
~~~~~~|\varrho|=1. \label{key} 
\end{equation}

This principle can be used to rederive Dirac and Weinberg equations \footnote{Actually the equations derived by Weinberg \cite{weinberg} 
missed the  $\varrho$ phase and are valid for the positive energy spinors 
only.}(see below). The question posed here is whether this 
principle can be generalized or not to other representations and 
used to derive new equations of motion (eom) for particles with 
$s\ge 1$. Alternatively it can give new insights into known eom's which 
could be rederived using this principle. 
The values of the phase $\varrho$ in Eq.(\ref{key}) can be restricted by 
imposing particular discrete
spacetime properties onto the spinors of interest,
like, say, parity covariance. Indeed, under parity the 
representations $(j,0)$ and $(0,j)$ are interchanged. 
At the quantum level this means that 
$\Pi\Phi _{R}\left( 0 \right)=\eta \Phi _{L}\left( 0 \right)$. Applying 
twice this operation and under the convention that we recover exactly the
 same state under two consecutive parity operations we obtain $\eta^2=1$.  
Transforming Eq.(\ref{key}) 
under parity \footnote{Strictly speaking this is intrinsic parity, i.e parity in the rest 
frame. When acting on $p$-dependent spinors we must also transform 
$\vec p\to -\vec p$ which in the rest frame is trivial. We will call i-parity to this 
transformation in the following to distinguish it for the full parity 
transformation.} we obtain
\begin{equation}
\Phi _{L}\left(\vec 0 \right) =\varrho ~\Phi _{R}\left(\vec 0 \right),  
\label{keytransf}
\end{equation}
\noi 
\noi which when  using Eq. (\ref{key}) yield $\varrho^2=1$ i.e 
$\varrho=\pm 1$. Furthermore, if we require parity as a good 
symmetry, we are forced to consider a space representation comprising both $(j,0)$ 
and $(0,j)$. The natural space is $(j,0)\oplus (0,j)$ or its companion 
$(0,j)\oplus (j,0)$. The corresponding spinors are 

\begin{equation}
\Psi_{RL} \left( \vec p \right)=\phi_R\left(\vec p \right)\oplus  
\phi_L\left(\vec p \right) =\left( \begin{array}{c}
\Phi _{R}\left( \vec p \right) \\ 
\Phi _{L}\left(\vec p\right)
\end{array}
\right), 
\end{equation}
and 
\begin{equation}
\Psi_{LR} \left(\vec p \right)=\phi_L\left( \vec p \right)\oplus  
\phi_R\left( \vec p \right) =\left( \begin{array}{c}
\Phi _{L}\left(\vec  p \right) \\ 
\Phi _{R}\left( \vec p \right)
\end{array}
\right), 
\end{equation}

These spinors transform under general Lorentz transformations as 
\begin{equation}
\Lambda_{RL} \left( \vec \theta ,\vec \varphi \right) \Psi_{RL} =\left( 
\begin{array}{cc}
e^{i\vec J\cdot (\vec \theta-i\vec \varphi )} & 0 \\ 
0 & e^{i\vec J\cdot (\vec \theta+i\vec \varphi )}
\end{array}
\right) \left( 
\begin{array}{c}
\Phi_R  \\ 
\Phi_L 
\end{array}
\right) \, ,
\end{equation}

\begin{equation}
\Lambda_{LR} \left( \vec \theta ,\vec \varphi \right) \Psi_{LR} =\left( 
\begin{array}{cc}
e^{i\vec J\cdot (\vec \theta+i\vec \varphi )} & 0 \\ 
0 & e^{i\vec J\cdot (\vec \theta-i\vec \varphi )}
\end{array}
\right) \left( 
\begin{array}{c}
\Phi_L  \\ 
\Phi_R 
\end{array}
\right)\, .
\end{equation}

Notice that under pure rotations the $\Psi_{RL}$ and the $\Psi_{LR}$ spinors 
transform identically which leads to a PI for the whole space which reads

\be
\Psi_{RL}(\vec 0)=\varrho \Psi_{LR}(\vec 0) ~{\rm with}~ \varrho=\pm 1. 
\label{key1}
\en

On the other hand, i-parity transformation for the fundamental representations 
$(j,0)$ and $(0,j)$ induces the following transformation for the $(j,0)\oplus (0,j)$ 
representation
\be
\Psi_{RL}(\vec p)\to \Pi\Psi_{RL}(\vec p)=\eta\Psi_{LR}(\vec p) 
~~ \Rightarrow ~~ \Pi =\eta\left( 
\begin{array}{cc}
0 & 1 \\ 
1 & 0
\end{array}
\right).
\label{iparity}
\end{equation}
\noi The factor $\eta=\pm 1$ is not relevant for the following 
derivation and will 
be omitted thorough this section. Using boost operators on the rest 
frame spinors 
\be
\Psi_{RL}(\vec p)= B_{RL}(\vec p)\Psi_{RL}(\vec 0)\, , \label{brl}
\en
with 
\be
B_{RL}(\vec p)= \left( \begin{array}{cc} 
B_R (\vec p) & 0 \\ 
0 & B_L (\vec p)
\end{array}\right)=\left( 
\begin{array}{cc}
e^{\vec J\cdot \vec\varphi} & 0 \\ 
0 & e^{-\vec J\cdot \vec\varphi}
\end{array} \right)\, ,
\en
with similar relations for $\Psi_{LR}$ and $B_{LR}(\vec p)$. 
Notice that $B_{RL}(-\vec p)=B_{LR}(\vec p)$. Furthermore, under i-parity 
\be
B_{RL}(\vec p)\to \Pi B_{RL}(\vec p)\Pi=  B_{LR}(\vec p)= 
B_{RL}(-\vec p)\, . 
\label{brlpar}
\en
Let us now boost the condition (\ref{key1})
\bq  \nonumber 
\Psi_{RL}(\vec p)&=& 
\varrho B_{RL}(\vec p)\Psi_{LR}(0)=
\varrho B_{RL}(\vec p)\Pi \Psi_{RL}(0) \\\nonumber
&=& \varrho B_{RL}(\vec p)\Pi 
B^{-1}_{RL}(\vec p)\Psi_{RL}(\vec p)\\ \nonumber
&=&\varrho B_{RL}(\vec p)\Pi 
B_{RL}(-\vec p)\Psi_{RL}(\vec p) \\ \nonumber
&=&\varrho B_{RL}(\vec p)\Pi^2 
B_{RL}(\vec p)\Pi \Psi_{RL}(\vec p) \\\nonumber 
&=&\varrho B^2_{RL}(\vec p) \Pi \Psi_{RL}(\vec p)\, ,
\eq

\noi where we have used Eqs. (\ref{iparity},\ref{brl},\ref{brlpar}) 
consecutively. This equation can be rewritten as 

\be
[B^2_{RL}(\vec p) \Pi - \varrho ] \Psi_{RL}(\vec p)=0,
\en

\noi or explicitly in terms of the angular momentum generators \footnote{If we keep 
the $\eta$ phase thorough, the $\varrho$ phase in Eq.(\ref{eomj}) 
must be replaced by $\eta\varrho =\pm 1$.}

\begin{equation}
\left( 
\begin{array}{cc}
-\varrho & ~ e^{2\vec J\cdot \vec\varphi}\\ 
 e^{-2\vec J\cdot \vec\varphi}& -\varrho
\end{array}
\right) \Psi_{RL} \left( \vec p\right) =0. \label{eomj}
\end{equation}

This is the equation of motion that any field in the $(j,0)\oplus (0,j)$ 
representation must satisfy. The explicit form of these equations in terms of
 momentum operators require to evaluate the exponentials. The explicit form - 
for arbitrary $j$ - of these exponentials can be obtained from the general 
relations for the hyperbolic functions given in appendix A of Ref.\cite{weinberg}. 
As an example, let us consider the cases $j=\frac{1}{2},1,\frac{3}{2}$.

\subsection{\,$\left( j=\frac{1}{2}\right) $: Dirac Equation.}

In the case $j=\frac{1}{2}$ , $J= \frac{\vec\sigma}{2}$ satisfy the 
{\it bilinear} algebra $ \lbrace J_i, J_j \rbrace =\frac{1}{2}\delta_{ij}$
which can be used to evaluate the exponential as  
\be
e^{\vec{\sigma }\cdot 
\vec{\varphi }}=~ \cosh \varphi ~ {\bf 1}~+~\left( \vec{\sigma }\cdot 
\widehat{p}\right) \sinh \varphi .\label{espin1/2}
\en
\noi From Eqs. (\ref{phi(E,p)},\ref{eomj},\ref{espin1/2}) we obtain 
\begin{equation}
\left( 
\begin{array}{cc}
-1 & \varrho ~ \frac{E+\vec{\sigma}\cdot \vec p }{m}\\ 
\varrho ~\frac{E-\vec{\sigma}\cdot \vec p }{m} & -1
\end{array}
\right) \Psi \left( \vec p \right) =0. \label{cuasieom1/2}  
\end{equation}
\noi In terms of the $4\times 4$ matrices
\begin{equation}
\gamma ^{0}=\left( 
\begin{array}{cc}
0 & 1 \\ 
1 & 0
\end{array}
\right) ,\qquad \gamma ^{i}=\left( 
\begin{array}{cc}
0 & -\sigma ^{i} \\ 
\sigma ^{i} & 0
\end{array}
\right)\, .  
\end{equation}
\noi We can rewrite Eq. (\ref{cuasieom1/2}) in Dirac's form

\begin{equation}
\left( \gamma ^{\mu }p_{\mu }-\varrho ~ m\right) \Psi 
\left( \vec p \right) =0. 
\label{diraceom}
\end{equation}
This is the conventional Dirac theory where $\varrho =1$ corresponds to 
positive energy spinors and $\varrho= -1$ to negative energy spinors. When
we switch to Dirac representation for the $\gamma$ matrices i-parity operator 
is diagonal and the corresponding spinors have opposite intrinsic parity. The 
particular assignment of intrinsic parity 
depends on the choice for the phase $\eta$ which remains arbitrary. 
It is worth 
noticing that  $\gamma_\mu$ satisfy Dirac algebra $\lbrace \gamma^\alpha , 
\gamma^\beta \rbrace = 2 g^{\alpha\beta}$ {\it which is just the covariant 
version of the algebra satisfied by $\frac{J_i}{s}$}.

\subsection{ \,$\left(\,  j=1\, \right)$.}

In the case $j=1$, the angular momentum operators, in addition to the Lie 
algebra of $SU(2)$, can be shown to satisfy also the {\it trilinear algebra}

\be
 J_i J_j J_k + J_k J_j J_i = \delta_{ij} J_k + \delta_{jk}J_i ,
\en

\noi which can be used to calculate the exponentials in Eq.(\ref{eomj}). 
Then, from 
Eq.(\ref{eomj}) we obtain
{\small 
\begin{equation}
\left( 
\begin{array}{cc}
-\varrho m^{2} \qquad\qquad  m^{2}+2E\left( \vec{J}\cdot \vec{p}\right)&
+2\left( \vec{J}\cdot \vec{p}\right) ^{2} \\ 
m^{2}-2E\left( \vec{J}\cdot \vec{p}\right) +2\left( 
\vec{J}\cdot \vec{p}\right) ^{2} & -\varrho m^{2}
\end{array}
\right) \left( 
\begin{array}{c}
\Phi _{R}\left( \vec{p}\right) \\ 
\Phi _{L}\left( \vec{p}\right)
\end{array}
\right) =0 . \label{eom1}
\end{equation}
}
\noi Defining now the matrices 

\begin{equation}
\gamma _{00}=\left( 
\begin{array}{cc}
0 & 1 \\ 
1 & 0
\end{array}
\right), \quad  
\gamma _{0i}=\gamma _{i0}=\left( 
\begin{array}{cc}
0 & J_{i} \\ 
-J_{i} & 0
\end{array}
\right), ~~~
\gamma _{ij}=g _{ij}\left( 
\begin{array}{cc}
0 & 1 \\ 
1 & 0
\end{array}
\right) +\left( 
\begin{array}{cc}
0 & \left\{ J_{i},J_{j}\right\} \\ 
\left\{ J_{i},J_{j}\right\} & 0
\end{array}
\right)\, ,
\end{equation}

\noi Eq.(\ref{eom1}) can be cast into Weinberg's form (up to the 
phase $\varrho$ which does not appear in Weinberg equations) 

\begin{equation}
\left[ \gamma _{\mu \nu }p^{\mu }p^{\nu }-\varrho~m^{2}\right] \Psi 
\left(\vec p\right)=0 . 
\end{equation}

\subsection{\, $\left(\, j=\frac{3}{2}\, \right) $. }

The generators for rotations in the case of $j=\frac{3}{2}$ can be shown to 
satisfy the {\it cuatrilinear algebra} 
\be
\{ J_{ij},J_{kl}\} = 5 (\delta_{ij} J_{kl} + \delta_{kl} J_{ij}) 
-\frac{9}{2}\delta_{ij}\delta_{kl}. \label{cuatrilinear}
\en
\noi where $ J_{ij}\equiv \{ J_i,J_j \} $. Similar calculations and  
defining the totaly symmetric matrices

\begin{equation}
\gamma _{000}=\left( 
\begin{array}{cc}
0 & 1 \\ 
1 & 0
\end{array}
\right),~~  
\gamma _{00i}=
\left( 
\begin{array}{cc}
0 & \frac{2}{3}J_{i} \\ 
-\frac{2}{3}J_{i} & 0
\end{array}
\right),~~~  
\gamma _{0ij}=
\left( 
\begin{array}{cc}
0 & \frac{1}{4} g_{ij}+\frac{1}{6}\left\{ J_{i},J_{j}\right\} \\ 
\frac{1}{4} g_{ij}+\frac{1}{6}\left\{ J_{i},J_{j}\right\} & 0
\end{array}
\right), 
\en
\be
\gamma _{ijk}=
\left( 
\begin{array}{cc}
0 & \frac{4}{3}J_{ijk}+\frac{7}{3}T_{ijk} \\ 
-\frac{4}{3}J_{ijk}-\frac{7}{3}T_{ijk} & 0
\end{array}
\right) \, ,
\end{equation}

\noi where

\begin{equation}
J_{ijk}=\frac{1}{6}\left( J_{i}\left\{ J_{j},J_{k}\right\} +J_{j}\left\{
J_{k},J_{i}\right\} +J_{k}\left\{ J_{i},J_{j}\right\} \right),~~  
T_{ijk}=\frac{1}{3}\left( g_{ij}J_{k}+g_{jk}J_{i}+ g_{ki}J_{j}\right)\, ,  
\end{equation}

\noi yield the equation of motion

\begin{equation}
\left[ \gamma _{\mu \nu \sigma }p^{\mu }p^{\nu }p^{\sigma }-
\varrho m^{3}\right]
\Psi \left(\vec p\right) =0 , 
\end{equation}

\noi which is an ${\cal O}(p^3)$ equation. 
These results can be easily generalized. For general 
$j$, the angular momentum operators, in addition to satisfy $SU(2)$ Lie 
algebra will also satisfy a $2j+1$-linear algebra which will yield  
\be
e^{2\vec J\cdot\varphi} \approx  {\cal O}(p^{2j})\, ,
\en
\noi and  fields transforming in the $(j,0)\oplus (0,j)$ representation will 
obey equations of the form 
\begin{equation}
\left[ \gamma ^{\mu _{1}\ldots \mu _{2j}}p_{\mu_1}\ldots p_{\mu_{2j}}-\varrho
m^{2j}\right] \Psi \left(\vec p \right) =0.  
\end{equation}

\noi i.e. equations of ${\cal O}(p^{2j})$ which are intractable for $j>1$.

\section{ Generalization to direct products.}

\subsection{{\rm $(j_1,0)\otimes (0,j_2)$} representation: Generalities.}

The simplest representations beyond  $(j,0)\oplus (0,j)$ are those 
obtained as tensor product of the
 $(j_1,0)$ and $(0,j_2)$ representations, i.e., the $(j_1,0)\otimes (0,j_2 )$ 
representations. The corresponding rotation operators are given by

\be
R(\vec\vartheta)=R_R(\vec\vartheta )\otimes R_L(\vec\vartheta )\equiv  
R_{RL}(\vec\vartheta )
=e^{i\vec J_1\cdot\vec\vartheta}
\otimes e^{i\vec J_2 \cdot\vec\vartheta}, \label{rotations} 
\en
\noi whereas for boosts we obtain
\be
B(\vec\varphi)=B_R(\vec\varphi )\otimes B_L(\vec\varphi )\equiv B_{RL}(\vec\varphi )
=e^{\vec J_1\cdot\vec\varphi}\otimes e^{-\vec J_2 \cdot\vec\varphi}, \label{boosts}
\en
 
The generators, in the tensor product basis (TPB), for these 
transformations satisfy (for arbitrary $\widehat n$)
\be
\vec J\cdot \widehat n =\frac{1}{i}\frac{\partial R_R(\vartheta)}
{\partial \vartheta} = \vec J_1\cdot\widehat n\otimes {\bf 1}_{(2j_2+1)
\times(2j_2+1)} +
{\bf 1}_{(2j_1+1)\times(2j_1+1)} \otimes\vec J_2\cdot\widehat n,
\en
\be
\vec K\cdot \widehat n =\frac{1}{i}\frac{\partial B_R(\varphi)}
{\partial \varphi} =\frac{1}{i} \vec J_1\cdot\widehat n
\otimes {\bf 1}_{(2j_2+1)
\times(2j_2+1)} -
{\bf 1}_{(2j_1+1)\times(2j_1+1)} \otimes\vec J_2\cdot\widehat n,
\en
\noi thus the generators for the $(j_1,0)\otimes (0,j_2)$ representation, 
in the TPB are 
\be
\vec J=\vec J_1\otimes {\bf 1} +
{\bf 1}\otimes\vec J_2,~~~
i\vec K =\vec J_1\otimes {\bf 1}- {\bf 1}\otimes\vec J_2,\label{genj1j2}
\en
\noi where we omitted the dimensionality of the unit matrices which 
can be easily traced.

Under i-parity the  $(j_1,0)\otimes (0,j_2 )$ 
representations are mapped onto the $(0,j_1)\otimes (j_2 ,0)$ 
representations which are unitarily equivalent to $(j_2,0 )\otimes (0,j_1)$  
(by unitarily equivalent here we mean the existence of a unitary 
transformation which connects these representations in the rest frame). 
Thus the 
construction of a parity invariant theory forces us to consider the 
$\left[(j_1,0)\otimes (0,j_2)\right]\oplus 
\left[(0,j_1)\otimes (j_2,0)\right]$ or $\left[(j_1,0)\otimes (0,j_2)\right]
\oplus \left[(j_2,0 )\otimes (0,j_1)\right]$  as our representation space, 
except in the case of $j_1=j_2$ where $(j_1,0)\otimes (0,j_1)$ spans 
an irreducible representation for the Parity operator. Indeed, under parity 
$(j_1,0)\otimes (0,j_1)$ goes into $(0,j_1)\otimes (j_1,0)$ which is 
unitarily equivalent to $(j_1,0)\otimes (0,j_1)$. 
 
The construction of the equations of motion for fields transforming as  
$\left[(j_1,0)\otimes (0,j_2)\right]$ is more transparent when we work 
with irreducible representations with respect to the rotations subgroup. 
Thus we need to pass from the tensor product basis  
(TPB) $|j_1,m_1\rangle \otimes |j_2,m_2\rangle $ to the total 
angular momentum basis (TAMB) $|j_1j_2;j,m\rangle $. 
The unitary transformation $U$ connecting these basis 
is the matrix whose elements are the corresponding Clebsch-Gordon 
coefficients $(\langle j_1m_1;j_2m_2|j_1j_2;jm\rangle $).  

Let us start with the explicit construction with the simplest 
case $j_1=j_2=\frac{1}{2}$.

\subsection{$(\frac{1}{2},0)\otimes (0,\frac{1}{2})$: Proca representation.}

The states corresponding to  $(\frac{1}{2},0)\otimes (0,\frac{1}{2})$ are 
$|\frac{1}{2}m\rangle_R\otimes |\frac{1}{2}m'\rangle_L $ whereas those 
belonging to  $(0,\frac{1}{2})\otimes (\frac{1}{2},0)$ are 
$|\frac{1}{2}m\rangle_L\otimes |\frac{1}{2}m'\rangle_R $.
Explicitly the spinors are given by
\be
\phi_{RL}\equiv \phi_R\otimes\phi_L =~l.c.~ \left(
\begin{array}{c}
|+ \rangle_R |+ \rangle_L\\
|+ \rangle_R |- \rangle_L\\
|- \rangle_R |+ \rangle_L\\
|- \rangle_R |- \rangle_L \\
\end{array}
\right),~~~
\phi_{LR}\equiv \phi_L\otimes\phi_R = ~l.c.~\left(
\begin{array}{c}
|+ \rangle_L |+ \rangle_R \\
|+ \rangle_L |- \rangle_R \\
|- \rangle_L |+ \rangle_R \\
|- \rangle_L |- \rangle_R \\
\end{array}
\right).
\en

\noi where {\it l.c.} stands for linear combination. 
In the latter equations we used the 
customary notation $|\frac{1}{2}\frac{1}{2}\rangle 
\equiv |+ \rangle$ and $|\frac{1}{2}-\frac{1}{2}\rangle \equiv |- \rangle$, 
and the $L,R$ subindices to denote their different transformation 
properties under boosts. Notice that $\phi_{LR} = U~\phi_{RL}$ with
\be
U\equiv \left(
\begin{array}{cccc}
1 & 0 & 0 & 0 \\
0 & 0 & 1 & 0 \\
0 & 1 & 0 & 0 \\
0 & 0 & 0 & 1 
\end{array}
\right)\, .
\en

Under i-parity 
\be
\phi_{RL}\longrightarrow \eta ~\phi_{LR}=\eta ~U \phi_{RL}
\Rightarrow \Pi =\eta~U\, ,
\en
where $\eta$ is a phase which is restricted to $\eta=\pm 1$ because $\Pi^2=1$.
The boost and rotation operators can be constructed as
\be
R_{RL}(\vec\theta)= e^{i\frac{\vec\sigma}{2}\cdot 
\vec\theta}\otimes e^{i\frac{\vec\sigma}{2}\cdot\vec\theta}
=R_{LR}(\vec\theta),~~~
B_{RL}(\vec\varphi)= e^{\frac{\vec\sigma}{2}\cdot\vec\varphi}\otimes 
e^{-\frac{\vec\sigma}{2}\cdot\vec\varphi},~~
B_{LR}(\vec\varphi)= e^{-\frac{\vec\sigma}{2}\cdot\vec\varphi}\otimes 
e^{\frac{\vec\sigma}{2}\cdot\vec\varphi}\, .
\en
\noi The corresponding generators read 
\be
\vec J_{RL}=\frac{1}{2}(\vec\sigma\otimes {\bf 1} + {\bf 1}\otimes\vec\sigma )
=\vec J_{LR}\, , 
~~~~ i\vec K_{RL}=\frac{1}{2}(\vec\sigma\otimes {\bf 1} - {\bf 1}
\otimes\vec\sigma )=-i\vec K_{LR}\, .
\en
\noi In terms of $E$ and $\vec p$ the boost operators for rest frame states read
\be
B_{RL}(\vec p)=\frac{1}{2m(E+m)}[E+m+\vec\sigma\cdot\vec p ]\otimes 
[E+m-\vec\sigma\cdot\vec p ]= B_{LR}(-\vec p)\, .
\en

It can be shown that under i-parity
\be
B_{RL}(\vec p)\longrightarrow \Pi B_{RL}(\vec p)\Pi = B_{RL}(-\vec p)=
B_{LR}(\vec p)\, .
\en

Notice that the PI $\phi_R(0)=\varrho \phi_L (0)$ 
and its i-parity transformed $\phi_L(0)=\varrho \phi_R (0)$ induces the 
following principle for the composed representation

\be
\phi_{RL}(0)=\phi_{LR}(0)\, .
\en
Boosting this equation and a little of algebra yields

\bq \nonumber
\phi_{RL}(\vec p)&=& 
 B_{RL}(\vec p)\phi_{LR}(0)=
 B_{RL}(\vec p)\Pi \phi_{RL}(0) \\\nonumber
&=&  B_{RL}(\vec p)\Pi 
B^{-1}_{RL}(\vec p)\phi_{RL}(\vec p)\\ \nonumber
&=& B_{RL}(\vec p)\Pi 
B_{RL}(-\vec p)\phi_{RL}(\vec p) \\ \nonumber
&=& B_{RL}(\vec p)\Pi^2 
B_{RL}(\vec p)\Pi \phi_{RL}(\vec p) \\\nonumber 
&=& B^2_{RL}(\vec p) \Pi \phi_{RL}(\vec p).
\eq

\noi Thus, the PI yields the following equation of motion in the TPB.
\be
[B^2_{RL}(\vec p) U - \eta ] \phi_{RL}(\vec p)=0.
\en

\noi Although we derived this equation for the case $j=\frac{1}{2}$ it is clear 
that it holds for any $j$ with the appropriate change in the generators of 
rotations. For $j=\frac{1}{2}$ the squared boost operator satisfy
\be
B^2_{RL}(\vec p)=B^2_R(\vec p)\otimes B^2_L(\vec p)=
\frac{1}{m^2}[E+\vec\sigma\cdot\vec p ]\otimes \left[ E-\vec\sigma\cdot\vec p 
\right],
\en
\noi and the equation of motion for the Proca representation, in the TPB, reads
\be
\left([E+\vec\sigma\cdot\vec p ]\otimes [E-\vec\sigma\cdot\vec p ] ~U~-\eta  m^2
\right)\phi_{RL}(\vec p)=0. \label{AKTPB}
\en

It is interesting to write this equation in the TAMB. This basis is related 
to the TPB as $|TAMB\rangle = M_{cb} |TPB\rangle$. Explicitly 
\be
\left(
\begin{array}{c}
|0,0\rangle \\ |1,1\rangle \\  |1,0\rangle\\ |1,-1\rangle
\end{array}
\right)
=\left(
\begin{array}{cccc}
0&\frac{1}{\sqrt{2}}&-\frac{1}{\sqrt{2}}&0\\
1&0&0&0\\
0&\frac{1}{\sqrt{2}}&\frac{1}{\sqrt{2}}&0\\
0&0&0&1 \\
\end{array}
\right)
\left(
\begin{array}{c}
|+\rangle_R |+\rangle_L \\ |+\rangle_R |-\rangle_L \\
|-\rangle_R |+\rangle_L \\ |-\rangle_R |-\rangle_L \\
\end{array}
\right). \label{cb}
\en

\noi Under this change of basis:
\be
\Pi \longrightarrow \tilde\Pi = M_{cb}\Pi M^{\dagger}_{cb} = 
\eta Diag(-1,1,1,1)\, ,
\en
\be
\vec J^{TPB}_{RL} \longrightarrow \vec J^{TAMB}_{RL}=
\left(
\begin{array}{cc}
\vec 0_{1\times 1} &\vec 0_{1\times 3}\\
\vec 0_{3\times 1}&\vec L\\
\end{array}
\right),~~
i\vec K^{TPB}_{RL} \longrightarrow i\vec K^{TAMB}_{RL}=
\left(
\begin{array}{cc}
\vec 0_{1\times 1} &\vec B^\dagger\\
\vec B &\vec 0_{3\times 3}\\
\end{array}
\right)\, ,
\en
\noi where  
\be
B^\dagger_1= \frac{1}{\sqrt{2}}(-1,0,1),~~
B^\dagger_2= -\frac{i}{\sqrt{2}}(1,0,1), 
~~B^\dagger_3=(0,1,0)\, ,  \label{Bi} 
\en
\noi and $\vec L$ denote the angular momentum operators for spin 1. 
These relations make clear that the field $\tilde\phi_{RL}\equiv M_{cb}\phi_{RL}$ 
describes a multiplet composed of a spin zero field and a spin 1 field with 
opposite intrinsic parities. The choice $\eta=-1$ yields 
$(\tilde\Pi)_{\mu\nu} = g_{\mu\nu}$ and with this choice $\tilde\phi_{RL}$ 
describes a multiplet of a spin 0 field with positive intrinsic parity and a 
spin 1 field with with negative intrinsic parity.

Transforming Eq.(\ref{AKTPB}) to the TAMB we obtain 
\be
[\tilde B^2_{RL}(\vec p) \tilde U + \eta  ] 
\tilde \phi_{RL}(\vec p)=0\, ,
\en
which yields 
\be
[{\cal O} - \eta m^2]\tilde\phi_{RL}(\vec p) = 0\, ,
\en
where
\be
{\cal O}=\left(
\begin{array}{cccc}
E^2+{\vec p}^2   & \sqrt{2} E p_+ & -2 E p_z            & -\sqrt{2} E p_- \\
 -\sqrt{2} E p_- & -(E^2-p^2_z)   & \sqrt{2} p_zp_+     & p^2_-            \\
2 E p_z          & \sqrt{2}p_zp_+ & -E^2-p^2_z + p_+p_- & -\sqrt{2}p_zp_- \\
\sqrt{2} E p_+   & p^2_+          & \sqrt{2}p_zp_+      & -(E^2-p^2_z)    \\
\end{array}
\right)\, ,
\en
or 
\be
( p^{\mu} p^{\nu}\Lambda_{\mu\nu}-\eta m^2 )
\tilde \phi (\vec p)=0 \, ,\label{1212}
\en
with
\be
\Lambda_{\mu\nu}=\Lambda_{\nu\mu},~~\Lambda_{00}= Diag (1,-1,-1,-1) ~~
{\rm etc.}
\en
This equation was first derived in Ref.~\cite{AK} using projectors 
techniques instead of a PI. In that work different (but equivalent) 
representation for the $\Lambda_{\mu\nu}$ matrices are obtained. 
An exhaustive analysis of the properties 
of this equation can also be found in Ref.~\cite{AK}. Here we just stress 
that this equation follows from the same principle as the Dirac and 
Weinberg equations.

\bi

\subsection{$(1,0)\oplus(\frac{1}{2},\frac{1}{2})\oplus(0,1)$: 
Kemmer-Duffin-Petieau representation.}

Historically some other formalisms for the description of particles with spin 
one have been considered. In particular the Kemmer-Duffin-Petieau (KDP) 
formalism \cite{KDP,marek} which uses the 
$(1,0)\oplus(\frac{1}{2},\frac{1}{2})\oplus(0,1)$ 
representation of the Lorentz group. Let us explore the possible consequences 
of the PI for this representation. 

The corresponding analysis is straightforward if we use the representation  
$(1,0))\oplus(0,1)\oplus(\frac{1}{2},\frac{1}{2})$ which is connected to the 
KDP representation by an obvious unitary transformation and will also be 
called KDP representation in the following. The corresponding spinors have the 
following structure
\be
\phi^{KDP}_{RL}=\left( \begin{array}{c}
\phi^{W}_{RL} \\ \phi^{P}_{RL} \\
\end{array} \right)\, ,
\en
where $\phi^{W}_{RL}$ and  $\phi^{P}_{RL}$ denote spinors in the 
$(1,0))\oplus(0,1)$ (Weinberg representation in the following) and 
Proca representations respectively. Under i-parity 

\bq \nonumber
\phi^{KDP}_{RL}=\phi^{W}_{RL}\oplus \phi^{P}_{RL} \longrightarrow 
\phi^{KDP}_{LR}&=&\phi^{W}_{LR}\oplus \phi^{P}_{LR}  
=(\Pi^{W}\phi^{W}_{LR})\oplus (\Pi^{P}\phi^{P}_{LR})\\ 
&=&\Pi^{KDP}\phi^{KDP}_{LR},
\eq
thus i-parity operator has the following structure
\be
\Pi^{KDP}\equiv \Pi^{W}\oplus \Pi^{P}= \left( \begin{array}{cc}
\Pi^{W}&0 \\ 0&\Pi^{P} \\
\end{array} \right).
\en
Clearly, there is no mixing between $\phi^{W}_{RL}$ and $\phi^{P}_{RL}$ under 
parity or Lorentz transformations.

\noi The PI for this representation reads
\be
\phi^{KDP}_{RL}(0)=\
\tilde\varrho \phi^{KDP}_{LR}(0)\, ,
\en
where now $\tilde\varrho$ stands for the block-diagonal matrix 
$Diag (\varrho {\bf 1}_{6\times 6},\eta {\bf 1}_{4\times 4})$. Boosting this 
equation we obtain 

\be
[{B^{KDP}(\vec p)}^2 \Pi^{KDP}-\tilde\varrho]~\phi^{KDP}_{RL}
(\vec p )=0\, ,
\en 
where ${B^{KDP}(\vec p)}^2= {B^{W}(\vec p)}^2\oplus {B^{P}(\vec p)}^2$, 
i.e. 
\be
\left[\left( \begin{array}{cc}
{B^{W}(\vec p)}^2 \Pi^{W}-\varrho & 0 \\
0 & {B^{P}(\vec p)}^2 \Pi^{P}-\eta \\
\end{array}
\right) \right]
 \left( \begin{array}{c}
\phi^{W}_{RL}(\vec p) \\ \phi^{P}_{RL}(\vec p) \\
\end{array} \right)=0.
\en

As a final result the equation splits into two independent equations, one for 
the Weinberg field  and another one for the Proca field with a common mass. 

\subsection{
$(\frac{1}{2},\frac{1}{2})\otimes [(\frac{1}{2},0)\oplus (0,\frac{1}{2})] $: 
Rarita-Schwinger representation.}

Let us now study the Rarita- Schwinger representation in the light 
of the PI. 
The corresponding spinors are constructed as a direct product of  the Dirac 
and Proca fields. Under i-parity the RS field transforms as follows

\be
\phi^{RS}_{RL}=\phi^{P}_{RL}\otimes \phi^{D}_{RL} \longrightarrow 
\phi^{RS}_{LR}=\phi^{P}_{LR}\otimes \phi^{D}_{LR}  
=(\Pi^{P}\phi^{P}_{LR})\otimes (\Pi^{D}\phi^{D}_{LR})=\Pi^{RS}\phi^{RS}_{LR}.
\en
with $\Pi^{RS}\equiv \Pi^{P}\otimes \Pi^{D}$. Rotation operators for the 
$RL\equiv \left[(\frac{1}{2},0)\otimes (0,\frac{1}{2})\right]\otimes 
\left[ (\frac{1}{2},0)\oplus (0,\frac{1}{2})\right] $ and the 
$LR\equiv \left[(0,\frac{1}{2})\otimes (\frac{1}{2},0)\right]\otimes 
\left[ (0,\frac{1}{2})\oplus (\frac{1}{2},0)\right] $ representations are
\be
R^{RS}_{RL}(\vec \vartheta )=R^{P}_{RL}(\vec \vartheta )\otimes R^{D}_{RL}
(\vec \vartheta ),~~~~~
R^{RS}_{LR}(\vec \vartheta )=R^{P}_{LR}(\vec \vartheta )\otimes R^{D}_{LR}
(\vec \vartheta ).
\en
\noi The identical transformation properties under rotations for the Dirac 
and Proca fields induces a PI for $\phi^{RS}_{LR}$ and $\phi^{RS}_{RL}$ in the 
rest frame. Boosts operators for the $RL$ representation reads
\be
B^{RS}_{RL}(\vec \varphi )=B^{P}_{RL}(\vec \varphi )\otimes B^{D}_{RL}(\vec 
\varphi ),
\en 
which under i-parity transforms as 
\be
\Pi^{RS} B^{RS}_{RL}(\vec p )\Pi^{RS} = B^{RS}_{RL}(-\vec p ). \label{bmp}
\en

The PI for this representation
\be
\phi^{RS}_{RL}(0)=\varrho \phi^{RS}_{LR}(0),
\en
\noi together with (\ref{bmp}) yields
\be
[{B^{RS}(\vec p)}^2 \Pi^{RS}-\varrho]~\phi^{RS}_{RL}(\vec p )=0.
\en 
Notice that 
\be
{B^{RS}(\vec p)}^2 ={B^{P}(\vec p)}^2\otimes {B^{D}(\vec p)}^2 =
{\cal O}(p^2)\otimes {\cal O}(p) = {\cal O}(p^3),
\en
and the PI yields an ${\cal O}(p^3)$ equation which we will not push further 
here due to the known acausalities for any ${\cal O}(p^3)$ equation.

\subsection{Other representations containing spin $\frac{3}{2}$}

To study other possibilities such as $(1,0)\otimes (0,\frac{1}{2})$ we need 
to consider in general the structure for the $(j_1,0)\otimes (0,j_2)$ 
representation with $j_1\neq j_2$. 
Under parity $(j_1,0)\otimes (0,j_2) \to  (0,j_1)\otimes (j_2,0)$. A theory 
for quantum fields which consider parity as a good symmetry would require to 
consider the whole $[(j_1,0)\otimes (0,j_2)] \oplus [(0,j_1)\otimes (j_2,0)]$
 space.  We denote  $(j_1,0)\otimes (0,j_2)$ as ``left'' 
representation and $ (0,j_1)\otimes (j_2,0)$ as ``right''representation 
thorough this section. In the TPB the generators for these representations read

\be
\vec J_L=\vec J_1\otimes {\bf 1} +
{\bf 1}\otimes\vec J_2,~~~
i\vec K_L  =-\vec J_1\otimes {\bf 1}+ {\bf 1}\otimes\vec J_2. \label{pgenj1j2L}
\en
\be
\vec J_R=\vec J_1\otimes {\bf 1} +
{\bf 1}\otimes\vec J_2,~~~
i\vec K_R  =\vec J_1\otimes {\bf 1}- {\bf 1}\otimes\vec J_2. \label{pgenj1j2R}
\en

Notice that the relations $J_L= J_R,~K_L= -K_R$, which are valid for the 
$(j,0)$ and $(0,j)$ representations, are also valid in this case. Hence, in 
the case at hand, when posed in the rest frame,  it is also impossible to 
distinguish fields transforming in the ``left'' from those transforming in 
the ``right'' representation which can be used to derive the corresponding 
equation of motion. As for the  parity operator it has 
in TPB a simple form that
follows from the transformation properties of the ``left'' and ``right'' 
representations. There, the generators for rotations and  boosts  
 for the whole $[(j_1,0)\otimes (0,j_2)] \oplus [(0,j_1)\otimes (j_2,0)]$ 
representation have a simple block-diagonal form

\be
\vec J = \left( 
\begin{array}{cc}
\vec J_R & 0 \\ 
0 & \vec J_L
\end{array}
\right),~~
\vec K = \left( 
\begin{array}{cc}
\vec  K_R & 0 \\ 
0 & \vec K_L
\end{array}
\right),
\en
\noi with $\vec J_L=\vec J_R$, $\vec K_L=-\vec K_R$ given by Eq.
(\ref{genj1j2}), whereas i-parity is represented by the operator 
\be
 \Pi = \left( 
\begin{array}{cc}
0 & {\bf 1} \\ 
{\bf 1} & 0
\end{array}\label{parity}
\right),
\en
and the corresponding equation of motion is
\be
[{B^{j_1j_2}(\vec p)}^2 \Pi-\varrho]~\phi_{RL}(\vec p )=0\, ,
\en 
which in general is an ${\cal O}(p^{2j_{max}})$ equation, 
with $j_{max}=j_1+j_2$ 
the maximum value of the total angular momentum. 
In particular for $j_1=1$ $j_2=\frac{1}{2}$ we obtain an 
${\cal O}(p^3)$ equation. 
We have studied many other possibilities for spin $\frac{3}{2}$ following 
this strategy. All of them yield ${\cal O}(p^3)$ equations.

Summarizing up to this point, although a PI can be formulated for the 
representations obtained as direct product of the simplest representations 
$(j,0)$ and $(0,j^\prime )$, the order of the corresponding eom's exhibit a 
clear pattern. The higher the spin, the higher the order of the corresponding 
eom is and for $s>1$ we obtain eom's ${\cal O}(p^3)$ or higher order in $p$. 
A complete search for the possible eom's that a field in a specific 
representation 
can satisfy requires to change our strategy. Indeed, using the PI we obtained 
equations of motion  which the corresponding free fields must necessarily 
satisfy. 
However, it is still possible that these free fields satisfy a different 
eom also. 
In the next section we use the information on the specific representation 
(the explicit form of  the generators of the HLG) in a different way and 
explore the possibilities for the existence of linear (Dirac-like) eom for 
that 
representation.

\section{Back to the basics: Covariance and linear equations.}

The order of the equation of motion for a field containing spin $j$, as 
dictated by the PI, can be understood from the algebra which, in addition to 
the Lie algebra, the generators of rotations satisfy. This additional algebra 
is different for different values of  $j$. Linearity of the eom for spin 
$\frac{1}{2}$ comes from the fact that 
generators in this case satisfy the {\it bilinear} algebra
\be
\lbrace J_i,J_j\rbrace = \frac{1}{2}\delta{ij} \label{alg12}
\en
whereas the order $p^2$ equation of motion for particles with $s=1$ (either 
Weinberg, Proca or A-K) comes from the {\it trilinear} algebra satisfied by the 
$s=1$ generators 
\be
 J_i J_j J_k + J_k J_j J_i = \delta_{ij} J_k + \delta_{jk}J_i , \label{alg1}
\en
and the order $p^3$ of the eom's containing spin $\frac{3}{2}$ comes from the 
{\it cuatrilinear} algebra which the $s=\frac{3}{2}$ generators fulfill
\be
\{ J_{ij},J_{kl}\} = 5 (\delta_{ij} J_{kl} + \delta_{kl} J_{ij}) 
-\frac{9}{2}\delta_{ij}\delta_{kl},
\label{alg32}
\en
\noi with $ J_{ij}\equiv \lbrace J_i,J_j \rbrace$.

The way out this pattern is the one followed by the KDP equation. 
This is an 
${\cal O} (p)$ equation for $ s=1$ fields. If we are able to rederive this 
equation from group theoretical arguments we will be on the road toward
the construction of linear equations for higher spin fields. 

The information on the representation we are working with is contained 
in the generators which can be explicitly constructed for a given 
representation. We can use this information and the constraints 
arising from Lorentz covariance to check if there exists or not a 
linear eom for a given representation. In this way we assume that 
the field $\psi$ in the given representation satisfy
\be
\left[ \beta_\mu p^\mu -m \right]\psi=0.
\en
Lorentz covariance requires $\beta_\mu $ to satisfy 
\be
[M_{\mu\nu}, \beta_{\alpha}]=i(g_{\nu\alpha}\beta_{\mu}-
g_{\mu\alpha}\beta_{\nu} )\, ,
\en
which in terms of rotations and boosts generators $J_i=\epsilon_{ijk}M^{jk}$ 
and $K_i=M_{0i}$ read
\be
[J_i, \beta_0]=0,~[J_i,\beta_j]=i\epsilon_{ijk}\beta_k ,~
[iK_i,\beta_0]=-\beta_i,~ [iK_i,\beta_j]=-\delta_{ij}\beta_0. \label{funcon}
\en
We use these relations to explicitly construct the matrices $\beta_\mu$. 

\subsection{\, $(j,0)\oplus (0,j)$: fields with single spin}

The generators for the $(j,0)\oplus (0,j)$ representation are 
\be
\vec J = \left( 
\begin{array}{cc}
\vec J_R & 0 \\ 
0 & J_L
\end{array}
\right),\qquad   
i\vec K = \left( 
\begin{array}{cc}
\vec J_R & 0 \\ 
0 & -J_L
\end{array}
\right), \label{genjj}  
\end{equation}
\noi where $\vec J_R=\vec J_L$ stands for the generators of rotations for a spin $j$ 
system. Let us write $\beta_0$ in the block-matrix form
\be
\beta_0 = \left( 
\begin{array}{cc}
b^0_{11} & b^0_{12} \\ 
b^0_{21} & b^0_{22}
\end{array}
\right),
\en
\noi where $b^0_{ij}$ are $2\times 2$ matrices. From the first of relations
(\ref{funcon}) we obtain 
\be
\left[\vec J^R,b^0_{ij}\right]=0.
\en
\noi Since $\vec J^R$ span an irrep of $SU(2)$, by Schur's lemma the 
$b^0_{ij}$ sub-matrices must be proportional to the identity 
$b^0_{ij}=b_{ij}{\bf 1}$, where $b_{ij}$ are numbers. Next we define 
$\beta_i$ by the third of relations (\ref{funcon})
\be
\beta_i \equiv [-iK_i,\beta_0]=2 \left( 
\begin{array}{cc}
0 & b_{12}J^R_i \\ 
-b_{21}J^R_i & 0
\end{array}
\right).
\en
The second of relations (\ref{funcon}) is satisfied by this matrix and 
an explicit calculation of the commutator in the fourth of relations 
(\ref{funcon}) yields 
\be
\left[K_i,\beta_j\right] =-2i \left( 
\begin{array}{cc}
0 & b_{12}\{J^R_i,J^R_j\} \\ 
 b_{21}\{J^R_i,J^R_j\} & 0
\end{array}
\right), 
\en
\noi and the fourth of relations(\ref{funcon}) requires $b_{11}=b_{22}=0$ and 
\be
\{J^R_i,J^R_j\}=\frac{1}{2}\delta_{ij}
\en

{\it The only representation $(j,0)\oplus (j,0)$ whose generators satisfy this 
relation is $j=\frac{1}{2}$}. For higher spin the only solution is the trivial 
one. Thus, we have shown that there exists no Dirac-like equation 
of motion for 
fields transforming in the $(j,0)\oplus (0,j)$ representation for
 $j>\frac{1}{2}$. 
Notice in pass that the most general form of Dirac matrices which is 
consistent with Lorentz covariance only, contains two arbitrary parameters
\be
\beta_0 = \left( 
\begin{array}{cc}
{\bf 0} & a {\bf 1} \\ 
b{\bf 1} & {\bf 0}
\end{array}\right),~~~
\beta_i = \left( 
\begin{array}{cc}
{\bf 0} & a\sigma_i \\ 
-b\sigma_i & {\bf 0}
\end{array}
\right)\, ,
\en
\noi where $a\equiv b_{12},~b\equiv b_{21}$. These two free parameters
 are just 
a consequence of the inequivalence of the $(j,0)$ and the $(0,j)$
 representation 
which are separately irreps of the HLG. If we require also invariance
 under parity 
we obtain $a=b$. If we further require that the equation describe a particle 
(anti-particle) of mass $m$ we need $a=1$ ($a=-1$). Under this circumstance 
the matrices $\beta_\mu$ satisfy the 
Dirac algebra $\lbrace \beta_\mu , \beta_\nu \rbrace = 2g_{\mu\nu}$ which 
looks 
like the ``covariantized''version of the algebra satisfied by
 $\frac{\vec J}{s}$ (see Eq.(\ref{alg12})).

\subsection{$(\frac{1}{2},\frac{1}{2})$: Proca representation}

Similar calculations for this representation shows that there exists 
no Dirac-like equation in this case.

\subsection{$(1,0)\oplus (\frac{1}{2},\frac{1}{2})\oplus (0,1)$ again: 
KDP Equation.}

The quantum states for KDP representation in the TPB and TAMB 
\be
\phi_{TPB}=~l.~c.\left(\begin{array}{c}
|1m_1\rangle_{R} \\ |\frac{1}{2} m\rangle_R \otimes |\frac{1}{2} m'\rangle_L \\
|1m_1\rangle_{L}\\
\end{array}
\right)\, , \qquad
\phi_{TAMB}=~l.~c.\left(\begin{array}{c}
|1m_1\rangle_{R} \\ 
|0,0\rangle \\
|1,m\rangle  \\
|1m_1\rangle_{L}\\
\end{array}
\right)\, ,
\en 

\noi are related by the unitary matrix ${\cal M}$ as: 
$\phi_{TAMB}= {\cal M}_{cb} \phi_{TPB}$ with
\be 
{\cal M}_{cb}=\left(\begin{array}{ccc}
{\bf 1}_{3\times 3}& & \\
 &M_{cb} & \\
 & & {\bf 1}_{3\times 3} \\
\end{array}
\right)\, ,
\en
\noi where $M_{cb}$ is given in Eq.(\ref{cb}). Under this change of basis 
the generators transform as 
\be
\vec J_{TPB}=\left(\begin{array}{ccc}
\vec L& & \\
 &\vec J^{(\frac{1}{2},\frac{1}{2})}_{TPB} & \\
 & & \vec L \\
\end{array}
\right) \longrightarrow
\vec J_{TAMB}=\left(\begin{array}{ccc}
\vec L& & \\
 &\begin{array}{cc} \vec 0 & \vec 0^{\dagger} \\ \vec 0& \vec L\\ 
\end{array}& \\
 & & \vec L \\
\end{array}
\right)\, ,
\en
where $\vec J^{(\frac{1}{2},\frac{1}{2})}_{TPB}=\frac{1}{2}(\vec \sigma 
\otimes {\bf 1} + {\bf 1}\otimes \vec \sigma )$

\be
i\vec K_{TPB}=\left(\begin{array}{ccc}
\vec L& & \\
 &i\vec K^{(\frac{1}{2},\frac{1}{2})}_{TPB} & \\
 & & -\vec L \\
\end{array}
\right) \longrightarrow
i\vec K_{TAMB}=\left(\begin{array}{ccc}
\vec L& & \\
 &\begin{array}{cc} \vec 0 & \vec B^{\dagger}\\ \vec B& \vec 0\\ 
\end{array}& \\
 & & -\vec L \\
\end{array}
\right)
\en
with $i\vec K^{(\frac{1}{2},\frac{1}{2})}_{TPB}=\frac{1}{2}(\vec \sigma 
\otimes {\bf 1} - {\bf 1}\otimes \vec \sigma )$ and $\vec B$ is given in 
Eq.(\ref{Bi}). After a straightforward calculation we obtain
\be
\beta_0= \left( \begin{array}{cccc}
0 &0& b_{13} & 0\\
0 &0 &0 &0 \\
b_{31} &0 &0 & b_{34} \\
0 &0& b_{43} &0 \\
\end{array}
\right)\, , ~~
\vec\beta= \left( \begin{array}{cccc}
\vec 0 & b_{13}\vec B & -b_{13} \vec L & \vec 0\\
-b_{31}\vec B^\dagger  &\vec 0 &\vec 0 & -b_{34}\vec B^\dagger \\
b_{31}\vec L &0 &0 &-b_{34}\vec L \\
0 & b_{43}\vec B & b_{43}\vec L &0 \\
\end{array}
\right)\, , \label{betaKDP}
\en
such that
\be
(\beta^{\mu}p_{\mu} -m )\phi_{TAMB}=0\, ,
\en
is covariant.  Notice that, similarly to the Dirac case, the most 
general equation consistent 
with Lorentz covariance only, contains four arbitrary parameters.
 Again this is just a consequence of the {\it four} inequivalent irreps of 
the HLG contained in KDP 
representation. The operator for i-parity can be directly obtained from the 
representation itself as
\be
\Pi= \eta \left( \begin{array}{cccc}
0 & 0 &0 & {\bf 1}_{3\times 3}\\
0 & 1 &0 & 0 \\
0 &0&-{\bf 1}_{3\times 3} & 0 \\
{\bf 1}_{3\times 3} & 0 &0 &0 \\
\end{array}
\right).
\en
\noi If we impose invariance under parity the four free parameters are 
related as:
 $b_{13}= -b_{43}$ and $b_{31}=-b_{34}$. Thus, we are left with two free
parameters which can be reduced to a $\pm \frac{1}{\sqrt{2}}$ factor 
if we require that all the fields contained in this equation have 
the same mass $m$. For these values of 
the parameters the matrices $\beta_\mu$ satisfy Kemmer algebra:
\be
\beta^\mu\beta^\nu\beta^\rho + 
\beta^\rho\beta^\nu\beta^\mu=g^{\mu\nu}\beta^\rho 
+ g^{\nu\rho}\beta^\mu .
\en
\noi
It is worth to remark that this algebra is just the 
``covariantized'' version of the algebra satisfied by $\frac{\vec J}{s}$ 
(see Eq.(\ref{alg1})).

The formal relation to Proca equation can be established if the components 
of the  KDP 
field are related to each other in a specific way. Indeed taking  
\be
\phi_{TPB}=\left(\begin{array}{c}
\vec E+i \vec B \\ 
A^{0} \\
\vec A \\
\vec E -i \vec B \\
\end{array}
\right)\, ,  
\en
where $\vec E^i \equiv G^{0i}$ $B^i \equiv \epsilon^{ijk}G_{jk} $ 
and the usual 
definition for the strength tensor $G_{\mu\nu}$, it can be easily 
shown that $A_\mu$ 
satisfy Proca equation. In other words, equivalence of the Proca and KDP 
equations 
require to use a very restricted class of fields in 
$(1,0)\oplus (\frac{1}{2},\frac{1}{2})\oplus (0,1)$, those fields in 
$(1,0)\oplus (0,1)$ constructed from $p^\mu A^{\nu}- p^\nu A^\mu $.

\subsection{Dirac-like Equation for the 
$\left[(1,0)\otimes (0,\frac{1}{2})\right]\oplus 
\left[(0,1)\otimes (\frac{1}{2},0)\right]$ representation.}

Let us denote through this subsection  
$\left[(1,0)\otimes (0,\frac{1}{2})\right]$ as ``left'' and 
$\left[(0,1)\otimes (\frac{1}{2},0)\right]$ as ``right'' representations.
The generators for the ``right'' representation 
 can be read from Eq. (\ref{genj1j2}) for the case 
$j_1=1,~~j_2=\frac{1}{2}$ as
\be
\vec J^{TPB}_R= \vec L\otimes {\bf 1}+{\bf 1}\otimes \vec S ,~~
i\vec K^{TPB}_R= \vec L\otimes {\bf 1}-{\bf 1}\otimes \vec S,
\en
\noi where we used the conventional $\vec L,\vec S$ notation for  
the  spin $1$ and spin$ \frac{1}{2}$ generators respectively. Now we transform 
everything from the tensor product basis $|1,m_l\rangle_r\otimes 
|\frac{1}{2},m_s\rangle_l $ \footnote{The labels $r,l$ remind us that the 
corresponding 
spinors belong to the $(1,0)$ and $(0,\frac{1}{2})$ respectively. 
This distinction 
is necessary since these spinors differ from spinors belonging to the 
 $(0,1)$ and $(\frac{1}{2},0)$ in their transformation properties 
under boosts.} to the TAMB $|1\frac{1}{2};jm\rangle$. 
Schematically
\be
|1\frac{1}{2};jm\rangle = U_R~|1,m_l\rangle_r\otimes |\frac{1}{2}m_s\rangle_l
\en
where $U_R$ is the matrix of Clebsch-Gordon coefficients
\be
U_R=\left(\begin{array}{cccccc}
1&0&0&0&0&0  \\
0&\sqrt{\frac{1}{3}}&\sqrt{\frac{2}{3}}&0&0&0 \\
0&0&0&\sqrt{\frac{2}{3}}&\sqrt{\frac{1}{3}}&0 \\
0&0&0&0&0&1 \\
0&\sqrt{\frac{2}{3}}&-\sqrt{\frac{1}{3}}&0&0&0 \\
0&0&0&\sqrt{\frac{1}{3}}&-\sqrt{\frac{2}{3}}&0 \\
\end{array}\right).\label{UR}
\en 
The Lorentz group generators are transformed accordingly to
\be
\vec J^{TAMB}_R=U_R\vec{J^{TPB}}_RU^{\dagger}_R=\left(\begin{array}{cc}
\vec{\cal J} & {\bf 0}_{4\times 2} \\
{\bf 0}_{2\times 4}&\vec S \\
\end{array}\right),~~~
i\vec K^{TAMB}_R=U_Ri\vec K^{TPB}_RU^{\dagger}_R=\left(\begin{array}{cc}
\frac{1}{3}\vec{\cal J} & \frac{2}{3}\vec B^\dagger \\
\frac{2}{3}\vec B& \frac{5}{3}\vec S \\
\end{array}\right)\, , \label{genr}
\en
\noi where $\vec{\cal J}$ stands for spin$\frac{3}{2}$ generators and
\be
B^\dagger_1\equiv \left(\begin{array}{cc}
-\sqrt{\frac{3}{2}} & 0 \\
0&- \sqrt{\frac{1}{2}}  \\
\sqrt{\frac{1}{2}}& 0  \\
0&\sqrt{\frac{3}{2}}\\
\end{array}\right),~~
B^\dagger_2\equiv i \left(\begin{array}{cc}
\sqrt{\frac{3}{2}} & 0 \\
0& \sqrt{\frac{1}{2}}  \\
\sqrt{\frac{1}{2}}& 0  \\
0&\sqrt{\frac{3}{2}}\\
\end{array}\right),~~
B^\dagger_3\equiv \left(\begin{array}{cc}
0 & 0 \\
\sqrt{2}& 0  \\
0 & \sqrt{2}  \\
0&0\\
\end{array}\right).
\en
The matrices $\vec B$ were obtained firstly in Ref.~\cite{hurley} by a different 
procedure and satisfy
\be
\begin{array}{cc}
{\cal J}_i{\cal J}_j + B^\dagger_i B_j =
i\frac{3}{2}\epsilon_{ijk}{\cal J}_k 
+\frac{9}{4}\delta_{ij}, ~~~~&
S_iS_j + B_i B^\dagger_j =-i\frac{3}{2}\epsilon_{ijk}S_k 
+\frac{9}{4}\delta_{ij}\, , \\
B_i{\cal J}_j-{\cal J}_jB_i=\frac{5}{2} i\epsilon_{ijk}B_k\, , &
S_iB_j-S_jB_i=-\frac{1}{2}i\epsilon_{ijk}B_k\, , \\
B_i{\cal J}_j-S_iB_j=\frac{3}{2}i\epsilon_{ijk}B_k\, , &
B_i{\cal J}_j-S_j B_i = i \epsilon_{ijk}B_k\, , \\
B^\dagger_iB_j-B^\dagger_jB_i=2i\epsilon_{ijk}{\cal J}_k\, , &
B_iB^\dagger_j-B_jB^\dagger_i=-4i\epsilon_{ijk}S_k\, . \\
\end{array}\label{relaciones}
\en
Let us now study the ``left'' representation. 
As discussed in the previous section, under parity 
$(1,0)\otimes (0,\frac{1}{2})\to (0,1)\otimes (\frac{1}{2},0)$, hence  parity 
has a simple representation in the TPB for the whole 
$\left[(1,0)\otimes (0,\frac{1}{2})\right]\oplus 
\left[(0,1)\otimes (\frac{1}{2},0)\right]$. The generators of Lorentz 
transformations in the TPB can be read from 
Eqs.~(\ref{pgenj1j2L},\ref{pgenj1j2R})
\be
\vec J^{TPB}_L=\vec L\otimes{\bf1}+{\bf 1}\otimes \vec S ,~~~
i\vec K^{TPB}_L=-\vec L\otimes{\bf1}+{\bf 1}\otimes \vec S .\label{genl}
\en
\noi Next we transform states to the TAMB 
\be
|1\frac{1}{2};jm\rangle_L = U_L~|1m_l\rangle_l\otimes|\frac{1}{2}m_s
\rangle_r\, ,
\en
where $U_L$ is the matrix of the corresponding Clebsch-Gordon coefficients
which has exactly the same form as $U_R$ in Eq.(\ref{UR}).
Transforming the Lorentz group generators 
for the left representation accordingly we obtain the simple result 
\be
\vec J^{TAMB}_L\equiv U_L\vec J_L U^\dagger_L = \vec J^{TAMB}_R,~~~
\vec K^{TAMB}_L\equiv U_L\vec K_L U^\dagger_L = -\vec K^{TAMB}_R. 
\label{genltamb}
\en
Finally, the change of the basis from the TPB to the TAMB for the complete 
$\left[(1,0)\otimes (0,\frac{1}{2})\right]\oplus \left[(0,1)\otimes 
(\frac{1}{2},0)\right]$ representation is accomplished 
by the unitary transformation 
\be
U=\left( \begin{array}{cc}
U_R & {\bf 0} \\
{\bf 0}& U_L \\
\end{array}\right).
\en
The generators for rotations and boosts for the 
complete $\left[(1,0)\otimes (0,\frac{1}{2})\right]\oplus 
\left[(0,1)\otimes (\frac{1}{2},0)\right]$
representation transform to
\be
\vec J^{TAMB}=\left( \begin{array}{cc}
\vec J^{TAMB}_R& {\bf 0} \\
{\bf 0}&\vec J^{TAMB}_L \\
\end{array}\right),~~~
\vec K^{TAMB}=\left( \begin{array}{cc}
\vec K^{TAMB}_R& {\bf 0} \\
{\bf 0}&\vec K^{TAMB}_L \\
\end{array}\right)\, , \label{gentotal}
\en
\noi where $\vec J^{TAMB}_R=\vec J^{TAMB}_L$ and $\vec K^{TAMB}_L = -
\vec K^{TAMB}_R $ are given in Eq.(\ref{genr}); 
whereas i-parity operator remains invariant
\be
\Pi^{TAMB} = \left( \begin{array}{cc}
{\bf 0}& {\bf 1} \\
{\bf 1}&{\bf 0} \\
\end{array}\right). \label{parirytamb}
\en
We have now all what we need to construct the linear equation of motion for 
fields in the  $\left[(1,0)\otimes (0,\frac{1}{2})\right]\oplus 
\left[(0,1)\otimes (\frac{1}{2},0)\right]$ representation which reads 
\be
\left(\beta^\mu p_\mu ~ - ~m \right)\Phi(\vec p)=0.
\en
\noi We now exploit constraints arising from Lorentz covariance in 
Eq.(\ref{funcon}). Firstly we write $\beta_0$ in the block-matrix form
\be
\beta_0 = \left( 
\begin{array}{cc}
b^0_{11} & b^0_{12} \\ 
b^0_{21} & b^0_{22}
\end{array}
\right),
\en
\noi where $b^0_{ij}$ are $6\times 6$ matrices. From the first of relations
(\ref{funcon}) we obtain 
\be
\left[\vec J^{TAMB}_R,b^0_{ij}\right]=0. \label{conmJb0ij}
\en
\noi The irreducibility of the generators $\vec{\cal J},\vec S $ in 
Eq.(\ref{genr}), severely restrict the most general form of the $b^0_{ij}$ 
matrices. Indeed, using Eq.(\ref{conmJb0ij}) Schur's lemma requires these 
matrices to have the form
\be
b^0_{11}=\left(\begin{array}{cc}
a_{11}{\bf 1}&{\bf 0} \\
 {\bf 0} & a_{22}{\bf 1} \\
\end{array}\right),
b^0_{12}=\left(\begin{array}{cc}
a_{13}{\bf 1} & {\bf 0} \\
 {\bf 0} & a_{24}{\bf 1} \\
\end{array}\right),
b^0_{21}=\left(\begin{array}{cc}
a_{31}{\bf 1} & {\bf 0} \\
 {\bf 0} & a_{42}{\bf 1} \\
\end{array}\right),
b^0_{22}=\left(\begin{array}{cc}
a_{33}{\bf 1} & {\bf 0} \\
 {\bf 0} & a_{44}{\bf 1} \\
\end{array}\right),
\en
\noi where $a_{ij}$ are constant.
The third of relations (\ref{funcon}) can be taken as the definition of 
$\beta_i$ and with this choice  the second of relations (\ref{funcon}) 
is automatically satisfied whereas the remaining  commutator, when using 
relations (\ref{relaciones}), requires
\be
a_{11}=a_{22}=a_{33}=a_{44}=0,~~~a_{24}=-\frac{1}{2} a_{13},~~~
a_{42}=-\frac{1}{2} a_{31}.
\en
\noi At the end the 
most general form for of the matrices $\beta_{\mu} $ 
consistent with Lorentz covariance depend on  two free 
parameters and have the following structure
\be
\beta^0=\left(\begin{array}{cc}
{\bf 0}& a~ A_0 \\
c~A_0 &{\bf 0}\\
\end{array}\right),~~ 
\beta^i=\frac{1}{3}\left(\begin{array}{cc}
{\bf 0}& -a~ A_i \\
c~A_i &{\bf 0}\\
\end{array}\right)\, ,
\en
\noi where $a\equiv a_{13},c\equiv a_{31}$ are arbitrary constant and 
\be
A^0=\left(\begin{array}{cc}
{\bf 1}& 0 \\
0 &-\frac{1}{2}{\bf 1}\\
\end{array}\right),~~ 
A^i=\left(\begin{array}{cc}
2{\cal J}_i&  B^\dagger_i \\
B_i & -5 S_i\\
\end{array}\right).
\en
An analysis of this equation in the rest frame shows that it 
describes a multiplet composed of a spin $\frac{3}{2}$ and a 
spin $\frac{1}{2}$ particle with masses $m$ 
and $\frac{m}{2}$ respectively. This equation was also obtained in 
 Ref.~\cite{sudarshan2} and firstly in Ref.~\cite{bhabha}. The 
matrices $\beta_\mu$ do not satisfy the single mass condition 
  \cite{harish,sudarshan2} for Dirac-like equations 
\be
(\beta \cdot p )^n=p^2(\beta\cdot p)^{n-2}~~~ {\rm for~some~integer~}~n. 
\label{singlemass}
\en 
\noi It is worth to remark that as a consequence of the 
Clifford algebra satisfied by the Dirac matrices, 
relation (\ref{singlemass}) holds in the case of Dirac for 
$n=2$, whereas the Kemmer algebra satisfied by the $\beta_\mu$ 
matrices in KDP theory 
ensure that the same relation holds, but in this case with $n=3$. Furthermore 
these algebras are just the ``covariant''
 version of the algebras satisfied by the operators $\frac{\vec J}{s}$. 
A generalization of these results to the case of spin $\frac{3}{2}$ points 
to the irreps of the covariant version of the spin $\frac{3}{2}$ cuatrilinear 
algebra satisfied by $\frac{\vec J}{s}$ (see Eq.(\ref{cuatrilinear})), namely 
\be 
\lbrace \beta_{\mu\nu},\beta_{\alpha\beta}\rbrace = \frac{20}{9} (g_{\mu\nu} 
\beta_{\alpha\beta} + g_{\alpha\beta} \beta_{\mu\nu}) 
-\frac{8}{9}g_{\mu\nu}g_{\alpha\beta}. \label{cuatri}
\en
\noi where $ \beta_{\mu\nu}\equiv \lbrace \beta_\mu,\beta_\nu \rbrace $. The 
possibility of a Dirac-like equation for fields with spin $\frac{3}{2}$ such 
that the corresponding matrices $\beta_\mu$ satisfy relation 
(\ref{singlemass}) for $n=4$ and also the cuatrilinear algebra in
 Eq.~(\ref{cuatri}) is presently under investigation.

\section{Summary and perspectives.}

In this work we briefly reviewed  the derivation of Dirac and Weinberg 
equations using  the ``principle of indistinguishability'' and 
explored its consequences for other irreducible representations of 
the Homogeneous Lorentz Group containing spin $\ge 1$. 
We obtained the following results:
i) For the representation $(\frac{1}{2},0)\otimes (0,\frac{1}{2})$ this 
principle yields a second order equation. 
The latter equation was originally obtained by 
Ahluwalia-Kirchbach (A-K) \cite{AK}. 
ii) The $(1,0)\oplus (\frac{1}{2},\frac{1}{2})\oplus (0,1)$ 
representation obeys a second order equation. 
The corresponding field can be decomposed into the direct sum 
of two fields with a common mass, one of them obeys the $j=1$ 
Weinberg equation 
and the other field satisfy the A-K equation.  
iii) An ${\cal O}(p^3)$ equation is derived from this principle for the 
$(\frac{1}{2},\frac{1}{2})\otimes [(\frac{1}{2},0)\oplus (0,\frac{1}{2})]$ 
representation. 
iv) All the explored representations containing spin $\frac{3}{2}$
 (and lower spins)  yield also ${\cal O}(p^3)$ equations. 

Changing our strategy we used the information on the representations 
in a different way. Exploiting the specific form of the generators of 
the HLG for a given representation we explored the  
existence of first order equations for different representations. 
We showed that there is no Dirac-like equation for  
$(j,0)\oplus(0,j)$ fields,  except for spin $\frac{1}{2}$. 
The corresponding matrices $\beta_\mu$ satisfy a Clifford algebra 
(the covariant 
version of the {\it bilinear} algebra satisfied by $\frac{\vec J}{s}$).
We also concluded that there is no linear equation  for the 
$(\frac{1}{2},\frac{1}{2})$ 
representation. As for the representation 
$ (1,0)\oplus (\frac{1}{2},\frac{1}{2})\oplus 
(0,1)$, it satisfies Kemmer-Duffin-Petieau  equation. The corresponding 
matrices $\beta_\mu$ satisfy Kemmer algebra which is just the covariant 
version of the {\it trilinear} algebra satisfied by the generators of 
rotations for $j=1$. 

There exists a Dirac-like equation for the $(1,\frac{1}{2})\oplus 
(\frac{1}{2},1)$ representation which describes a multiplet of two 
particles, one 
with spin $\frac{3}{2}$ and mass $m$ and another with spin $\frac{1}{2}$ and 
 mass $\frac{m}{2}$. Based on the single mass condition for Dirac-like 
equations and on the results previously described we conjectured 
existence of a Dirac-like 
equation for spin $\frac{3}{2}$ fields such that the 
corresponding matrices satisfy 
the single mass condition Eq.(\ref{singlemass}) for $n=4$ and the cuatrilinear 
algebra in Eq.(\ref{cuatri}).

\section{Acknowledgments}

It is my pleasure to thank D.V. Ahluwalia and M. Kirchbach for their useful 
comments and remarks during the presentation of this work at Zacatecas Forum in 
Physics 2002. I also thank M. G. Carrillo Ruiz  and M. Nowakowski for an 
initial collaboration on this topic. Work supported by CONACYT under grant 
458100-5-37234-E.

\end{document}